\documentclass[aps,preprint,amsmath,showpacs]{revtex4}
\begin{document}

\title{ANALYTIC STUDY OF PERSISTENT CURRENT IN A TWO-CHANNEL DISORDERED MESOSCOPIC RING}
\author{Jean Heinrichs}
\email{J.Heinrichs@ulg.ac.be} \affiliation{D\'{e}partement de
Physique, B{5a}, Universit\'{e} de Li\`{e}ge, Sart Tilman, B-{4000}
Li\`{e}ge, Belgium}
\date{\today}

\begin{abstract}
We present an extensive analytical study of persistent current in a weakly disordered two-chain cylindrical ring threaded by an Aharonov-Bohm flux $0 < \phi <\phi_0/2$ (with $\phi_0$ the flux quantum) and described by the Anderson model. The effect of the disorder reveals a strong reduction of the persistent current for flux values near $\phi_0/4$.

In conjunction with the pure system (zeroth order) current profile averaged over numbers of electrons and earlier results for the effect of disorder in one-dimensional rings, our two-channel results provide a simple interpretation of salient features of numerical results of Bouchiat and Montambaux (BM) for persistent current in an assembly of many-channel disordered rings.  Single-channel (one-dimensional) effects are responsible for the dip in the persistent obtained by BM near $\phi=0$ and the corresponding peak near $\phi_0/2$, while the effect of disorder in independent channel pairs accounts for abrupt decreases of current superimposed to a continuous linear decay as the flux value $\phi_0/4$ is approached from above and from below, respectively.  The persistent current in the two-channel ring involves a free particle current averaged over electron numbers of periodicity $\phi_0/2$, and a dominant disorder effect which has periodicity $\phi_0$.
\end{abstract}


\maketitle

\section{INTRODUCTION}\label{secI}

An Aharonov-Bohm flux threading a metallic or semiconducting ring leads to a persistent equilibrium current even if the ring is disordered, provided it is of mesoscopic size smaller than the mean free path for inelastic scattering \cite{1}.  This is a consequence of the quantum coherence of the electronic wavefunction, which is not destroyed by elastic impurity scattering.  Intensive study of persistent current was initiated many years ago with the appearance of the seminal paper by B\"{u}ttiker, Imry and Landauer\cite{2} predicting persistent current in a one-dimensional disordered ring.  The persistent current is periodic in the magnetic flux $\phi$ with a period expressed in terms of the flux quantum $\phi_0=hc/e$ (with $h$ the Planck constant, $c$ the speed of light and $-e$ the electron charge).   The periodicity of the persistent current is due to the modification of the periodic boundary condition for the wavefunction $\psi(x)$ along the ring of length $L$ in the presence of the flux $\phi$, namely

\begin{equation}\label{eq1}
\psi(x+L)=\psi(x)e^{i\frac{2\pi\phi}{\phi_0}}\quad .
\end{equation}•
The persistent current in a particular energy level $E_n$ is related to the corresponding flux derivative by

\begin{subequations}\label{eq2}
\begin{align}
I_n
&=-c\frac{\partial E_n}{\partial\phi}\label{eq2a}\quad ,\\
\intertext{where $c$ is the speed of light.  The total persistent current in the ring is}\nonumber\\
I 
&= \sum_n I_n \quad .\label{eq2b}
\end{align}
\end{subequations}•

The access to quantum-mechanical features of solids renders accurate measurements of persistent current highly desirable.  An important step in this direction has been achieved recently with the appearance of high-precision measurements of persistent current by two different groups\cite{3,4}, showing for the first time quantitative agreement with theoretical models for non-interacting electrons in diffusive rings (see also \cite{5,6}).

The work of B\"{u}ttiker et al.\cite{2} for a one-dimensional ring was later amplified and extended by detailed analytical studies\cite{7,8} of the effect of disorder on persistent current, using the Anderson tight-binding model with random atomic site energies.  We focus, in particular, on the analysis of \cite{8} (referred to as I in the following) which leads to a convenient exact perturbation theory for the energy levels of the ring, which will be generalized below in the case of a disordered two-channel ring.  Persistent current has been discussed in \cite{7,8} in various flux intervals close to $\phi=0$. {\em Throughout this paper we consider specifically the interval $0<\phi <\phi_0/2$} in which extensive numerical simulations of persistent current have been performed by Bouchiat and Montambaux (BM)\cite{9} for ten-channel disordered rings.  An important objective of our work is to interpret the simulation results\cite{9} by adapting the results of \cite{8} for a one-dimensional (one-channel) ring and new results for persistent current in a two-channel disordered ring (the thinnest multi-channel ring!) whose derivation forms the core of the present article.  The ring-model studied numerically in \cite{9} is a quasi-one-dimensional Anderson model with on-site disorder and nearest-neighbour inter- as well as intra-chain hopping.  We focus attention on the fig. 4 of \cite{9} which reveals, in particular, the existence of an important dip in the persistent current near $\phi=0$ and a corresponding peak near $\phi_0/2$.  Abrupt current decays are also observed in the figure when approaching the value $\phi_0/4$ from below and from above, respectively.  Another specific feature of the fig. 4 in \cite{9} is that the persistent current averaged over even- and odd numbers of electrons \cite{10} has periodicity $\phi_0/2$.  An important observation of BM is that the shape of their fig. 4 is independent of the number of channels, $M$.  This suggests indeed that important features of the figure might be observed already in single- and two-channel systems as is confirmed in Sect.\ref{secIVb} below, by adapting, in particular the results of I (averaging over even- and odd numbers of electrons).

The special interest in studying transport in disordered quasi-one-dimensional mesoscopic rings involving a large number $M$, of parallel channels, is the existence in these systems of a wide domain of metallic conduction extending over length scales between the elastic mean free path $\ell$ and the localization length $\xi=M\ell$, for metallic conduction\cite{12}.  The recent appearance of accurate persistent current measurements in diffusive electron systems\cite{3,4} has been followed by a renewed interest in theoretical studies of such systems\cite{21,23}.  In particular, Ref \cite{21} addresses issues such as ensemble averaging and interchannel correlated disorder which are related in a broad sense to aspects of this paper.

The paper is organized as follows.  In Sect.\ref{secII} we introduce the double-chain Anderson model for the disordered ring and derive an exact eigenvalue equation in terms of the transfer matrix in the simple case of pairwise identical random site energies on the two chains.  In Sect.\ref{secIII} we develop a second order perturbation theory for the two-channel ring eigenvalues.  In Sect.\ref{secIVa} we obtain the dominant flux-dependence of the persistent current from the eigenvalues averaged over the disorder and in \ref{secIVb} we compare our results with the fig. 4 of \cite{9} and discuss the periodicity of the persistent current.  Some additional  remarks are presented in \ref{secV}.

\section{TWO-CHAIN CYLINDRICAL RING MODEL AND EIGENVALUE EQUATION}\label{secII}

We consider a vertical cylindrical strip threaded by a flux $\phi$, constituted by two superposed circular tight-binding chains (1 and 2) described by the Anderson model.  The system obeys the usual tight-binding equations (with $\alpha=\frac{2\pi\phi}{N\phi_0}$)

\begin{align}\setcounter{equation}{2}
-e^{i\alpha} \varphi^1_{n+1}-e^{-i\alpha}\varphi^1 _{n-1}
&=(E-\varepsilon_{1n})\varphi^1_n-h \;\varphi^2_n
\label{eq3}\quad ,\\
-e^{i\alpha} \varphi^2_{n+1}-e^{-i\alpha}\varphi^2_{n-1}
&=(E-\varepsilon_{2n})\varphi^2_n-h \;\varphi^1_n, n=2,3,\ldots, N-1\quad ,
\label{eq4}\\
-e^{i\alpha} \varphi^1_{2}-e^{-i\alpha}\varphi^1 _{N}
&=(E-\varepsilon_{11})\varphi^1_1-h \;\varphi^2_1\quad ,
\label{eq5}\\
-e^{i\alpha} \varphi^1_{1}-e^{-i\alpha}\varphi^1 _{N-1}
&=(E-\varepsilon_{1N})\varphi^1_N-h \;\varphi^2_N\quad ,
\label{eq6}\\
-e^{i\alpha} \varphi^2_{2}-e^{-i\alpha}\varphi^2 _{N}
&=(E-\varepsilon_{21})\varphi^2_1-h \;\varphi^1_1\quad ,
\label{eq7}\\
-e^{i\alpha} \varphi^2_{1}-e^{-i\alpha}\varphi^2 _{N-1}
&=(E-\varepsilon_{2N})\varphi^2_N-h \;\varphi^1_N\quad ,
\label{eq8}
\end{align}•
involving $N$ one-orbital atomic sites, $n=1,2,\ldots N$ of spacing $a$ per chain\cite{7,8}.  Here $\varphi^{i}_{n}$ denotes the amplitude of an eigenstate wavefunction at site $n$ on chain $i$ ($i=1,2$), $\varepsilon_{in}$ and $E$ denote the atomic energy of site $n$ on chain $i$ and energy eigenvalues in units of minus a constant nearest-neighbour hopping parameter.  Finally, $h$ stands for the ratio of the hopping parameter between a given site on one chain to its vertical nearest-neighbour on the other chain and, minus a constant parameter of nearest-neighbour hopping on the individual chains.

We now transform the Eqs. (\ref{eq3}-\ref{eq8}) for the two-chain cylindrical strip (wire) by defining a basis of independent channels for wave transmission in the absence of disorder.  These channels are defined by amplitude bases

\begin{equation}\label{eq9}
\begin{pmatrix}
\vdots\\
\psi^i_n\\
\vdots
\end{pmatrix}•=
\hat U^{-1}
\begin{pmatrix}
\vdots\\
\varphi^i_n\\
\vdots
\end{pmatrix}•
\quad ,
\end{equation}•
in which the interchain hopping terms in (\ref{eq3}-\ref{eq8}) are diagonal

\begin{equation}\label{eq10}
\begin{pmatrix}
\psi^1_n\\\psi^2_n
\end{pmatrix}•=
\widehat U
\begin{pmatrix}
\varphi^1_n\\\varphi^2_n
\end{pmatrix}\quad ,\quad
\widehat U\frac{1}{\sqrt 2}
\begin{pmatrix}
1 & 1 \\  1 & -1
\end{pmatrix}\quad .
\end{equation} 
By transforming Eqs (\ref{eq3}-\ref{eq8}) in the channel bases \eqref{eq10} we obtain successively

\begin{align}
-\begin{pmatrix}\label{eq11}
e^{i\alpha}\psi^1_{n+1}+e^{-i\alpha}\psi^1_{n-1}\\
e^{i\alpha}\psi^2_{n+1}+e^{-i\alpha}\psi^2_{n-1}
\end{pmatrix}•
& =
\begin{pmatrix}
E-h-\frac{1}{2}(\varepsilon_{1n}+\varepsilon_{2n})
&
\frac{1}{2}(\varepsilon_{2n}-\varepsilon_{1n})\\
\frac{1}{2}(\varepsilon_{2n}-\varepsilon_{1n})
&
E+h-\frac{1}{2}(\varepsilon_{1n}+\varepsilon_{2n})
\end{pmatrix}•
\begin{pmatrix}
\psi^1_n\\
\psi^2_n
\end{pmatrix}•
\quad ,\\
-\begin{pmatrix}\label{eq12}
e^{i\alpha}\psi^1_{2}+e^{-i\alpha}\psi^1_{N}\\
e^{i\alpha}\psi^2_{2}+e^{-i\alpha}\psi^2_{N}
\end{pmatrix}•
& =
\begin{pmatrix}
E-h-\frac{1}{2}(\varepsilon_{11}+\varepsilon_{21})
&
\frac{1}{2}(\varepsilon_{21}-\varepsilon_{11})\\
\frac{1}{2}(\varepsilon_{21}-\varepsilon_{11})
&
E+h-\frac{1}{2}(\varepsilon_{11}+\varepsilon_{1N})
\end{pmatrix}•
\begin{pmatrix}
\psi^1_1\\
\psi^2_1
\end{pmatrix}
\quad ,\\
-\begin{pmatrix}
e^{i\alpha}\psi^1_{1}+e^{-i\alpha}\psi^1_{N-1}\\
e^{i\alpha}\psi^2_{1}+e^{-i\alpha}\psi^2_{N-1}
\end{pmatrix}•
& =
\begin{pmatrix}\label{eq13}
E-h-\frac{1}{2}(\varepsilon_{1N}+\varepsilon_{2N})
&
\frac{1}{2}(\varepsilon_{2N}-\varepsilon_{1N})\\
\frac{1}{2}(\varepsilon_{2N}-\varepsilon_{1N})
&
E+h-\frac{1}{2}(\varepsilon_{1N}+\varepsilon_{2N})
\end{pmatrix}•
\begin{pmatrix}
\psi^1_N\\
\psi^2_N
\end{pmatrix}
\quad ,
\end{align}
The secular equation for the eigenvalues of the two-chain cylindrical ring are obtained from the boundary condition which follows from rewriting Eqs (\ref{eq11}-\ref{eq13}) in terms of transfer matrices $\widehat P_n$ defined by

\begin{equation}\label{eq14}
\begin{pmatrix}
\psi^1_{n+1}\\
\psi^1_{n}\\
\psi^2_{n+1}\\
\psi^2_{n}
\end{pmatrix}•
=
e^{-i\alpha}\widehat P_n
\begin{pmatrix}
\psi^1_{n}\\
\psi^1_{n-1}\\
\psi^2_{n}\\
\psi^2_{n-1}
\end{pmatrix}•
,n=2,3,\ldots N-1\quad ,
\end{equation}•

\begin{equation}\label{eq15}
\begin{pmatrix}
\psi^1_{2}\\
\psi^1_{1}\\
\psi^2_{2}\\
\psi^2_{1}
\end{pmatrix}•
=
e^{-i\alpha}\widehat P_1
\begin{pmatrix}
\psi^1_{1}\\
\psi^1_{N}\\
\psi^2_{1}\\
\psi^2_{N}
\end{pmatrix},
\begin{pmatrix}
\psi^1_{1}\\
\psi^1_{N}\\
\psi^2_{1}\\
\psi^2_{N}
\end{pmatrix}•=
e^{-i\alpha}\widehat P_N
\begin{pmatrix}
\psi^1_{N}\\
\psi^1_{N-1}\\
\psi^2_{N}\\
\psi^2_{N-1}
\end{pmatrix}\quad .
\end{equation}•
The compatibility of the results of iterating \eqref{eq14} with the boundary conditions \eqref{eq15} readily yields the secular equation for the energy eigenvalues of the ring:

\begin{equation}\label{eq16}
\det \biggl[\widehat{\openone} -e^{-i \frac{2\pi\phi}{\phi_0}}\prod^{N}_{n=1}\widehat P_n\biggr]=0\quad ,
\end{equation}•
where

\begin{equation}\label{eq17}
\widehat{P}_n=
\begin{pmatrix}
-(E-h)+ \frac{1}{2}(\varepsilon_{1n}+\varepsilon_{2n})
& -e^{-i\alpha}
& -\frac{1}{2}(\varepsilon_{1n}-\varepsilon_{2n})
& 0\\
e^{i\alpha}
& 0
& 0
& 0\\
-\frac{1}{2}(\varepsilon_{2n}-\varepsilon_{1n})
& 0
&-(E+h)+ \frac{1}{2}(\varepsilon_{1n}+\varepsilon_{2n})
&-e^{i\alpha}\\
0
&0
&e^{i\alpha}
&0
\end{pmatrix}•\quad .
\end{equation}

While the solution of the eigenvalue equation (\ref{eq16}-\ref{eq17}) required for the study of persistent currents in the cylindrical ring is generally complicated, it simplifies considerably by assuming a certain short-range correlation between the site energies on one chain and those on the other chain.  Thus, in addition to assuming the random site energies $\varepsilon_{1m}$ and $\varepsilon_{2m}$ ($m=1,2,\ldots N$) to be uncorrelated i.e.

\begin{equation}\label{eq18}
\langle\varepsilon_{pm}\varepsilon_{qn}\rangle
=\varepsilon^2_{0} \delta_{m,n}\delta_{p,q}\quad ,
\end{equation}•
as in the usual Anderson model, we impose here that the energy of a site on one chain and the energy of the adjacent nearest neighbour on the other chain coincide in any realization i.e.

\begin{equation}\label{eq19}
\varepsilon_{1m}=\varepsilon_{2m}\equiv \varepsilon_{m}, m=1,2,\ldots N\quad .
\end{equation}•
We recall that models with correlated disorder became popular in the context of localization in linear chain systems when Phillips and coworkers\cite{13} first showed that they lead to the existence of delocalized electron states in one-dimensional disordered systems.  In particular, Sedrakyan and Ossipov\cite{14} have recently studied delocalized states in a disordered two-chain ladder model with a interchain correlation which is analogous to \eqref{eq19}.  For completeness sake we mention that the effect of short-range disorder correlation on persistent current has been studied recently by us in one dimension, using a simple generalization of the Anderson model\cite{15} in which one defines a subset of pairs of nearest-neighbour sites whose energies are correlated while the energies of all other pairs are uncorrelated.

With the correlation \eqref{eq19} the transfer matrix $\widehat P_n$ reduces to the simplier block-diagonalized form

\begin{equation}\label{eq20}
\widehat P_n=\widehat P_{1n}\oplus\widehat P_{2n}\quad ,
\end{equation}•
where
\begin{equation}\tag{20a}
\widehat P_{1n}=
\begin{pmatrix}
-(E-h)+\varepsilon_n
& -e^{-i\alpha}\\
e^{i \alpha}
&0
\end{pmatrix}•\quad ,
\widehat P_{2n}=
\begin{pmatrix}
-(E+h)+\varepsilon_n
& -e^{-i\alpha}\\
e^{i \alpha}
&0
\end{pmatrix}•\quad ,
\end{equation}•\
with $\varepsilon_{m}\equiv\varepsilon_{1m}=\varepsilon_{2m}, m=1,2,\ldots N$ denoting the common value of the random energies of adjacent nearest-neighbour sites along the ring.
Similarly, under \eqref{eq19}, the ring transfer matrix

\begin{equation}\label{eq21}
\widehat R=\prod^{N}_{n=1}\widehat P_n\quad ,
\end{equation}•
with matrix elements $R_{ij}$, reduces to the block-diagonal form

\begin{equation}\label{eq22}
\widehat R= \widehat R_1\oplus\widehat R_2\quad ,
\end{equation}•
with
\begin{equation}\tag{22a}
\widehat R_1=
\begin{pmatrix}
R_{11} & R_{12} \\ R_{21} & R_{22}
\end{pmatrix}•,
\widehat R_2=
\begin{pmatrix}
R_{33} & R_{34} \\ R_{43} & R_{44}
\end{pmatrix}\quad ,
\end{equation}•
where $\det \widehat R_1=\det\widehat R_2=1$ since $\widehat R_1$ and $\widehat R_2$ are products of unimodular matrices.

Using the results (\ref{eq20}-20a), \eqref{eq21} and  (\ref{eq22}-22a) the equation \eqref{eq16} factorizes in the form

\begin{equation}\label{eq23}
\biggl[1-e^{-i\frac{2\pi\phi}{\phi_0}}(R_{11}+R_{22})+e^{-i\frac{4\pi\phi}{\phi_0}}\det\widehat R_1\biggr]
\biggl[1-e^{-i\frac{2\pi\phi}{\phi_0}}(R_{33}+R_{44})+e^{-i\frac{4\pi\phi}{\phi_0}}\det\widehat R_2\biggr]=0\quad ,
\end{equation}•
which reduces to the following final equation for the eigenvalues of the ring in terms of partial traces of the transfer matrix

\begin{equation}\label{eq24}
\biggl[R_{11}+R_{22}-2\cos\frac{2\pi\phi}{\phi_0}\biggr]
\biggl[R_{33}+R_{44}-2\cos\frac{2\pi\phi}{\phi_0}\biggr]=0\quad .
\end{equation}•

We conclude this section by emphasizing that the ring transfer matrix \eqref{eq22} defined by block-diagonal matrices \eqref{eq17} (for $\varepsilon_{1n}=\varepsilon_{2n}$) constitutes a microscopic realization of the transfer matrix of independent conducting channels as envisionned in the phenomenological picture of a multichannel wire composed of independent conducting channels.  In any case the solution of the eigenvalue equation \eqref{eq24} is considerably simplier analytically than the solution of \eqref{eq16} for uncorrelated site energies.

\section{WEAK DISORDER PERTURBATION THEORY OF ENERGY EIGENVALUES}\label{secIII}

The energy eigenvalues from which the persistent current may be obtained are related to the eigenvalues of the transfer matrix (\ref{eq22}-22a).  In order to study the effect of a weak disorder by a perturbation theory we first determine the eigenvalues in the absence of disorder.  In this case 

\begin{equation}\label{eq25}
\widehat R=\widehat P^N,\quad\text{with }\widehat P=\widehat Q_1\oplus\widehat Q_2\quad ,
\end{equation}•
where
\begin{equation}\tag{25a}
\widehat Q_1=
\begin{pmatrix}
-(E-h)&-e^{-i\alpha}\\
e^{i\alpha}& 0
\end{pmatrix}•,
\widehat Q_2=
\begin{pmatrix}
-(E+h)&-e^{-i\alpha}\\
e^{i\alpha}& 0
\end{pmatrix}\quad ,
\end{equation}•
In order to obtain the elements of $\widehat P^N$ we diagonalize $\widehat P$ by means of a similarity transformation.  The eigenvalues of $\widehat P$ are

\begin{subequations}\label{eq26}
\begin{align}
\lambda^1_\pm
&=\frac{1}{2}
\biggl[-(E-h)\pm\sqrt{(E-h)^2-4}\biggr] \quad,\label{eq26a}\\
\lambda^2_\pm
&=\frac{1}{2}
\biggl[-(E+h)\pm\sqrt{(E+h)^2-4}\biggr]\quad , \label{eq26b}
\end{align}•
\end{subequations}
which correspond to unperturbed energy bands in terms of wavenumber variables $s_1,s_2$ defined by

\begin{subequations}\label{eq27}
\begin{align}
\lambda_\pm^1
&=e^{\pm is_1}\label{eq27a}\\
\lambda_\pm^2
&=e^{\pm is_2}\quad ,\label{eq27b}
\end{align}•
\end{subequations}•
with
\begin{subequations}\label{eq28}
\begin{align}
E\equiv E_h(s_1)=h-2\cos s_1,\label{eq28a}\\
E\equiv E_{-h}(s_2)=-h-2\cos s_2\label{eq28b}
\end{align}•
\end{subequations}•
where we assume $h\geq 1$ and refer to the $s_1$- and $s_2$-band as the upper and the lower band, respectively.  The similarity transformation which diagonalizes $\widehat P$ in terms of the eigenvalues \eqref{eq27} is defined by the eigenvector matrix

\begin{equation}\label{eq29}
\widehat U=U_1\oplus\widehat U_2
\end{equation}•
with

\begin{equation}\tag{29a}
\widehat U_i=
\begin{pmatrix}
e^{i(s_i-\alpha)}&e^{-i(s_i+\alpha)}\\
1&1
\end{pmatrix}•, i=1,2\quad .
\end{equation}•
Using the relation $\widehat P^m=\widehat U(\widehat U^{-1}\widehat P\widehat U)^m\widehat U^{-1}$ we finally obtain, from (\ref{eq27a}-\ref{eq27b}) and (\ref{eq29},29a),

\begin{equation}\label{eq30}
\widehat P^m=\widehat Q_1^m\oplus\widehat Q^m_2
\end{equation}•		
with
\begin{equation}\tag{30a}	
\widehat Q_i^m=\frac{1}{\sin s_i}
\begin{pmatrix}
\sin(m+1)s_i & -\sin m s_i e^{-i\alpha}\\
\sin m s_i e^{i\alpha} & -\sin(m-1)s_i
\end{pmatrix}•, i=1,2\quad .
\end{equation}•
In the absence of disorder we insert \eqref{eq25}, \eqref{eq30} and (30a) in \eqref{eq24} and obtain for the wavenumbers in the energy bands (\ref{eq28a}-\ref{eq28b})

\begin{subequations}\label{eq31}	
\begin{align}
s_1\equiv s_1(k)
&=\frac{2\pi}{N}\biggl(k+\frac{\phi}{\phi_0}\biggr),
k=1,2,\ldots,N	\quad , \label{eq31a}\\
s_2\equiv s_2(k')
&=\frac{2\pi}{N}\biggl(k'+\frac{\phi}{\phi_0}\biggr),
k'=1,2,\ldots,N\quad . \label{eq31b}
\end{align}•
\end{subequations}•

Next we develop a perturbation theory to discuss the effect of a weak site-energy disorder on the eigenvalues of the two-channel ring threaded by a flux $\phi$.  This generalizes our earlier treatment\cite{8} of the effect of disorder on persistent current in a one-dimensional ring.  The latter\cite{8} has also been applied recently for studying the effect of short-range correlated disorder on persistent current in one dimension\cite{15}.

We now wish to solve \eqref{eq24} for weak random fluctuations of the site energies.  To this end we expand the transfer matrix to successive orders in the $\varepsilon_i$ in \eqref{eq19} and obtain the corresponding first and second order corrections for the perturbed energy levels of the (coupled) two-chain system of the form

\begin{equation}\label{eq32}
E\equiv E_{s_1s_2}(k,k')=E^{(0)}_{s_1s_2}+E^{(1)}_{s_1s_2}+E^{(2)}_{s_1s_2}+\ldots, k, k'=0,\pm 1,\pm 2\ldots\quad ,
\end{equation}•		
where
\begin{equation}\tag{32a}	
E^{(0)}_{s_1s_2}(k,k')=-2(\cos s_1+\cos s_2)\quad .
\end{equation}•
The transfer matrices $\widehat P_n$ in \eqref{eq17} are split into unperturbed and perturbed parts\cite{8}

\begin{equation}\label{eq33}
\widehat P_n=\widehat P+\widehat V_n=
\widehat P -\biggl[\Delta E-\frac{1}{2}(\varepsilon_{1n}+\varepsilon_{2n})\biggr]
\hat e\oplus\hat e,\hat e=
\begin{pmatrix}
1&0\\0&0
\end{pmatrix}\quad ,•
\end{equation}•
where
\begin{equation}\label{eq34}
\Delta E=E^{(1)}_{s_1s_2}+E^{(2)}_{s_1s_2}+\ldots\quad ,
\end{equation}•
and the matrix product $\prod^{N}_{n=1}\widehat P_n$ is expanded to quadratic order in the quantities
$\biggl[\Delta E-\frac{1}{2}(\varepsilon_{1n}+\varepsilon_{2n})\biggr]$.  This yields\cite{8}

\begin{equation}\label{eq35}
\begin{split}
\widehat R=\prod^{N}_{n=1}\widehat P_n=\widehat P^N+\sum^N_{m=1}\widehat P^{m-1}\widehat V_m\widehat P^{N-m}\\
+\sum^N_{n=2}\sum^{n-1}_{m=1}\widehat P^{m-1}\widehat V_m\widehat P^{n-m-1}\widehat V_n\widehat P^{N-n}+\ldots\quad .
\end{split}•
\end{equation}•

Note that the expansion in powers of $\widehat V_n$ in \eqref{eq35} goes beyond a systematic expansion to successive orders in the site energy perturbation since e.g. the terms linear in $\widehat V_n$, besides being linear in $\varepsilon_n$, depend on all orders of the energy level perturbation via $\Delta E$.  Nevertheless \eqref{eq35} leads to a systematic determination of the successive order perturbations of the energy levels since the solution of the eigenvalue equation \eqref{eq24} imposes that the disorder effects in the partial traces of the transfer matrix must vanish.  The evaluation of the first and second order corrections in the energy levels \eqref{eq32} proceeds by performing the matrix products entering in \eqref{eq35}, in order to obtain the diagonal elements of the ring transfer matrix in the partial traces in \eqref{eq24}.

The contribution of the linear correction term in \eqref{eq35} to the trace of $\widehat R$ is given by

\begin{equation}\label{eq36}
\text{Tr}\widehat R=
\biggl(\frac{1}{\sin s_1}+\frac{1}{\sin s_2}\biggr)
\sin\frac{2\pi\phi}{\phi_0}\sum^N_{m=1}
\biggl[E_{s_1s_2}^{(1)}-\frac{1}{2}(\varepsilon_{1m}+\varepsilon_{2m})\biggr]
+\text{O}(\varepsilon^2)\quad ,
\end{equation}
using \eqref{eq30}, (30a) and \eqref{eq33}.  Since according to \eqref{eq24} eigenvalues of the disordered two-channel  ring correspond to values of the partial traces which are independent of the disorder, it follows from \eqref{eq36} that the first order effect in the eigenvalues in the correlated site energies model (\ref{eq18}-\ref{eq19}) is

\begin{equation}\label{eq37}
E_{s_1s_2}^{(1)}=\frac{1}{2N}\sum_n (\varepsilon_{1m}+\varepsilon_{2m})\quad ,
\end{equation}•
except for flux values equal to integer multiples of $\phi_0/2$.  Note that these flux values are precisely those at which the free particle spectrum of the ring is degenerate\cite{7}.

At quadratic order in the site energies, on the other hand, there are two types of contributions in Eq. \eqref{eq35}, $\widehat R\equiv \widehat R'+\widehat R''$, the first one

\begin{equation}\label{eq38}
\widehat R'\equiv -E^{(2)}_{s_1s_2}\sum^N_{m=1}\widehat P^{m-1}
(\hat e\oplus\hat e)\widehat P^{N-m}\quad ,
\end{equation}•
and the second one

\begin{equation}\label{eq39}
\widehat R''\equiv \sum^N_{n=2}\sum^{n-1}_{m=1}
(E^{(1)}_{s_1s_2}-\varepsilon_n)(E^{(1)}_{s_1s_2}-\varepsilon_m)
\widehat P^{m-1}
(\hat e\oplus\hat e)
\widehat P^{n-m-1}
(\hat e\oplus\hat e)
\widehat P^{N-n}\quad .
\end{equation}•
The explicit evaluation $\text{Tr}\widehat R'$, using \eqref{eq30} and (30a), yields

\begin{equation}\label{eq40}
\text{Tr}\widehat R'=E^{(2)}_{s_1s_2}
\biggl[\frac{1}{\sin^2 s_1}
\biggl(N\cos s_1\cos N s_1-\frac{3}{4}\frac{\sin N s_1}{\sin^3 s_1}\biggr)
+\text{same with }s_1\rightarrow s_2\biggr]\quad .
\end{equation}•
Similarly, the evaluation of $\text{Tr}\widehat R''$ leads to

\begin{equation}\label{eq41}
\text{Tr}\widehat R''=
\sum^N_{n=2}\sum^{n-1}_{m=1}(E^{(1)}_{s_1 s_2}-\varepsilon_m)(E^{(1)}_{s_1s_2}-\varepsilon_n)
\biggl[\frac{\sin (n-m)s_1\sin(N+m-n)s_1}{\sin^2 s_1}
+\text{same with }s_1\rightarrow s_2\biggr]
\quad .
\end{equation}•
The condition $\text{Tr}\widehat R'+\text{Tr}\widehat R''=0$ which follows from \eqref{eq24} thus yields the final result

\begin{equation}\label{eq42}
\begin{split}
E_{s_1s_2}^{(2)}(k,k')=
-\biggl[
\frac{1}{\sin^2 s_1}
\biggl(
N\cos s_1\cos N s_1-\frac{3}{4}\frac{\sin N s_1}{\sin^3 s_1}
\biggr)
+\text{same with }s_1\rightarrow s_2
\biggr]^{-1}\\
+\sum^N_{n=2}\sum^{n-1}_{m=1}
(E^{(1)}_{s_1s_2}-\varepsilon_m)(E^{(1)}_{s_1s_2}-\varepsilon_n)
\biggl[\frac{\sin (n-m)s_1\sin (N+m-n)s_1}{\sin^2 s_1}
+\text{same with }s_1\rightarrow s_2
\biggr] \quad ,
\end{split}•
\end{equation}•
which will be exploited in the following section for discussing the effect of the disorder on the persistent current.

\section{EFFECT OF DISORDER ON THE PERSISTENT CURRENT}\label{secIV}
\subsection{Calculation of the persistent current}\label{secIVa}

In this subsection, like in our previous work\cite{8,15} we discuss the explicit form of the persistent current obtained simply from the energy levels of the ring averaged over the disorder.  Indeed, the more accurate approach which consists in finding the persistent current (using (2a) from Eq. \eqref{eq42} before averaging over the disorder leads to results whose further analysis is intractable analytically.  By averaging \eqref{eq42}, over the disorder, using (\ref{eq18}-\ref{eq19}), we get, after some simple algebra,

\begin{equation}\label{eq43}
\langle E_{s_1s_2}^{(2)}(k,k')\rangle=
-\frac{\varepsilon^2_0}{4}
\frac{\biggl(\frac{(N-1)\cos 2Ns_1}{\sin^2 s_1}-\frac{\sin(N-1)s_1}{\sin^3 s_1}\biggr)+\text{same with }s_1\rightarrow s_2}
{\biggl(\frac{N\cos s_1\cos Ns_1}{\sin^2 s_1}-\frac{3}{4}\frac{\sin N s_1}{\sin^5s_1}\biggr)+\text{same with }s_1\rightarrow s_2}\quad ,
\end{equation}•		
which reduces, for large $N$, to the form

\begin{equation}\tag{43a}	
\langle E_{s_1s_2}^{(2)}\rangle=
-\frac{\varepsilon^2_0}{4}
\Biggl(\frac{\frac{\cos 2N s_1}{\sin^2 s_1}+\text{same with }s_1\rightarrow s_2}
{\frac{\cos s_1\cos N s_1}{\sin^2 s_1}+\text{same with }s_1\rightarrow s_2}
\Biggr)\quad ,
\end{equation}•
up to corrections of order $1/N$. Using (\ref{eq31a}-\ref{eq31b}) this expression may be rewritten exactly in terms of an important fixed flux-dependent prefactor which is common to all energy  levels $\langle E_{s_1s_2}(k,k')\rangle$ of the ring:
 
\begin{equation}\label{eq44}
\langle E_{s_1s_2}^{(2)}(k,k')\rangle=\frac{\varepsilon^2_0}{4}
\biggl(
\frac{1}{\cos\frac{2\pi\phi}{\phi_0}}-2\cos \frac{2\pi\phi}{\phi_0}
\biggr)
\frac{\sin^2 s_1+\sin^2 s_2}{\cos s_1\sin^2 s_2+\cos s_2\sin^2 s_1}\quad .
\end{equation}•		
Note that this expression is not valid for flux values equal to odd integer multiples of $\phi_0/4$ at which the perturbation theory clearly breaks down.

The effect of the disorder on the persistent current of an electron in an energy level \eqref{eq32} of the ring defined approximately from the averaged second order energy perturbation, namely

\begin{equation}\label{eq45}
I^{(2)}_{s_1s_2}=-c\frac{\partial \langle E^{(2)}_{s_1s_2}\rangle}{\partial\phi}\quad ,
\end{equation}•		
is given by
\begin{equation}\label{eq46}
I^{(2)}_{s_1s_2}(k,k',\phi)=
-\varepsilon^2_0\frac{\pi e}{2h}
\biggl(\sin\frac{2\pi\phi}{\phi_0}\biggr)
\biggl(2+\frac{1}{\cos^2\frac{2\pi\phi}{\phi_0}}\biggr)
f_{s_1s_2}+\text{O}\biggl(\frac{1}{N}\biggr)\quad ,
\end{equation}•		
with

\begin{equation}\tag{46a}	
f_{s_1s_2}(k,k')=
\frac{\sin^2 s_1+\sin^2 s_2}{\cos s_1\sin^2 s_2+\cos s_2\sin^2 s_1}\quad,
\end{equation}•
whose differentiation in \eqref{eq45} yields the terms of order $1/N$ which have been ignored in \eqref{eq46}.
Clearly, for $\phi/\phi_0\ll N/2\pi$ the coefficient of $f_{s_1s_2}$ in \eqref{eq46} represents the dominant flux dependence of the disorder contribution, $I^{(2)}=\sum_{k,k'}I^{(2)}_{s_1s_2}(k,k',\phi)$ in the total persistent current in the two-channel system\cite{16}.  In particular, this coefficient reveals a strong decrease of the persistent current due to the disorder for fluxes approaching the value $\phi_0/4$ from either side (the value $\phi_0/4$ itself lying obviously outside the domain of validity of the perturbation theory).
The explicit (numerical) evaluation of $I^{(2)}$, say in the typical case of a half-filled upper energy-band $s_1$ and a completely filled lower band $s_2$ lies beyond the framework of our analytic treatment and will not be pursued further.

\subsection{Persistent current periodicity, single- and two-channel effects in simulation results for a multi-channel ring\cite{9}}\label{secIVb}

There exist two fundamental periods in the persistent current problem, namely the nearest-neighbour hopping period $N\phi_0$ in the hamiltonian of Eqs. (\ref{eq3}-\ref{eq8}) and the revolution period $\phi_0$ for an electron on the ring subjected to the flux $\phi$, which affects the boundary condition \eqref{eq1}.  The periodicity $N\phi_0$ is responsible for the period $\phi_0/2$ of the free particle persistent current averaged over even and odd numbers of electrons in a one-dimensional ring\cite{7,8} as recalled below.  On the other hand, the periodicity $\phi_0/2$ is also obtained exactly for the effect of the disorder in a one-dimensional ring.  Since, however, the $\phi_0/2$ periodicity of the averaged free particle persistent current was not explicitely mentioned in \cite{8}, we briefly discuss it here, along with the influence of the disorder in one dimension.
These results are important for  our qualitative interpretation below of salient features of numerical simulation results of Bouchiat and Montambaux (fig. 4 of \cite{9}) for the effect of weak disorder on persistent current in a ten-channel (cylindrical) ring, in terms of single- and two-channel contributions obtained in Ref. \cite{8} and in the present paper, Eqs (\ref{eq46}, 46a).

Let
\begin{equation}\label{eq47}
I^0\equiv\langle I^0_{Ne}\rangle=\frac{1}{2}
\biggl(I^0_{\text{even}}+I^0_{\text{odd}}\biggr)\quad ,
\end{equation}•		
be the persistent current in the unperturbed half-filled upper tight-binding band in (32a), averaged over even and odd numbers of electrons $Ne$\cite{10}.  The currents 
$I^0_{\text{even}}$ and $I^0_{\text{odd}}$ are given by \cite{7,8}

\begin{subequations}\label{eq48}
\begin{align}
I^0_{\text{even}}
&\simeq -I_0\biggl(\frac{2\phi}{\phi_0}-1\biggr)\quad ,\\
I^0_{\text{odd}}
&\simeq -I_0\frac{2\phi}{\phi_0} ,0<\frac{\phi}{\phi_0}<\frac{1}{2}\quad ,
\end{align}•
\end{subequations}
and are illustrated by the figures 3(a) and 3(b) of \cite{7} (whose legends should be inverted \cite{15}).  Their determination by summing contributions from the  occupied energy levels below the Fermi level is easily visualized from the energy level spectrum as a function of flux shown in fig. 2 of \cite{7}.  The current in \eqref{eq47} is thus

\begin{equation}\label{eq49}
I^0=-I_0\biggl(\frac{2\phi}{\phi_0}-\frac{1}{2}\biggr), 0<\frac{\phi}{\phi_0}<\frac{1}{2}\quad ,
\end{equation}•		
whose periodicity is $\frac{\phi_0}{2}$.  Here $I_0=\frac{ev_F}{L}$ is the persistent current of an electron at the Fermi level and $v_F=\frac{2\pi}{\hbar}\frac{Ne}{N}$ is the Fermi velocity.  Next we recall the effect of a weak disorder on the magnitude of the persistent current in a one-dimensional ring given by Eqs (\ref{eq49}-\ref{eq50}) of I\cite{8}.  After averaging the effects obtained in \cite{8} for even and for odd numbers of electrons as in \eqref{eq49}, we obtain the total persistent current in the disordered half-filled one-dimensional ring:

\begin{equation}\label{eq50}
I=-I_0
\biggl(\frac{2\phi}{\phi_0}-\frac{1}{2}\biggr)
\biggl[1-\frac{N\varepsilon^2_0}{8\sin^2(2\pi\phi/\phi_0)}\biggr],
0<\frac{\phi}{\phi_0}<\frac{1}{2}\quad ,
\end{equation}•		
which also has periodicity $\phi_0/2$.  We note, incidentally, that \eqref{eq50} provides a simple qualitative interpretation of the numerical results of Trivedi and Browne\cite{11} for persistent current in one-dimensional disordered rings averaged over even and odd numbers of electrons (fig. 4 of \cite{11} and fig. 8 of \cite{9}).

At present we wish to compare our results for persistent current averaged over electron numbers in the single- and two-channel rings with dominant features shown by the flux-dependence of similarly averaged persistent current in a ten-channel ring in fig. 4 of Bouchiat and Montambaux\cite{9}.  The 10-channel disordered ring is described by the Anderson-model using a rectangular distribution of the site energies of width $W=0.2$ (in units of the nearest-neighbour hopping parameter)\cite{17} and the persistent current is averaged over numbers of electrons $150\leq N_e\leq 250$ in the range $0<\phi <\phi_0/2$.

The pertinence of a comparison of our results for one- and two-channel systems with the multichannels results of fig. 4 of \cite{9} is due to the fact that the latter were found to be independent of the number of channels\cite{9}.  On the other hand, the inclusion of results for the two-channel system besides those for a one-dimensional (single-channel) ring for the comparison with the fig. 4 of \cite{9} enriches our model through the novel effects related to the interchain hopping.

We now come to the detailed comparison of our perturbation results for persistent currents in one- and two-channel disordered rings with the numerical results for a ten-channel system in fig. 4 of Ref. \cite{7}.
First we consider the free particle persistent current \eqref{eq49} which shows, in particular, the period halving effect resulting from averaging the current over even and odd numbers of electrons in fig. 3 of \cite{4}.  This expression provides clearly a good "zeroth order" approximation of the overall current profile in fig. 4 in \cite{9}.  Next, the disorder effect in the persistent current \eqref{eq50} averaged over numbers of electrons, for a one-dimensional ring accounts qualitatively for the relatively large current dip near $\phi=0$ and a corresponding peak near $\phi=\phi_0/2$ due to disorder, in fig. 4 of \cite{9}.  Finally, the disorder contribution $I^{(2)}=\sum_{k,k'}I^{(2)}_{s_1s_2}$ in the persistent current in the two-channel ring involves the dominant fixed flux-dependent overall factor in \eqref{eq46}, which leads to a strong decrease with respect to the uniform decay \eqref{eq49} for $\phi\rightarrow\phi_0/4$ from below and to a similar strong decrease with respect to the uniform increase of current for $\phi\rightarrow\phi_0/4$ from above.  These current reductions are in line with a corresponding strong reduction of current with respect to a uniform decrease in fig. 4 of \cite{9} (which passes through zero at $\phi=\phi_0/4$) for $\phi\rightarrow\phi_0/4$ from below, starting from $\phi=0,216\;\phi_0$ up to $\phi=0,233\;\phi_0$, and with a corresponding reduction of current with respect to a uniform increase of unperturbed current in fig. 4 of \cite{9} for $\phi\rightarrow\phi_0/4$ from above, starting from the value $\phi\simeq 0,275\;\phi_0$ down to $\phi\simeq 0,254\;\phi_0$.  Finally, in our reference to the fig. 4 of \cite{9} we have excluded the domain $0,233\;\phi\leq\phi\leq 0,254\;\phi_0$ from our discussion not just for convenience but mainly because the perturbation theory breaks down at $\phi=0.25\;\phi_0$.  On the other hand, our identification of the most significant features of the persistent current in multi-channel system with single-channel and coupled two-channel disorder effects suggests the possibility of identifying further finer features of the fig. 4 of Bouchiat and Montambaux\cite{9} with 3-4, ...-channel effects.  This might suggest an analogy between the persistent current problem in multi-channel systems and e.g. the interactions in a typical many-body system such as a molecular crystal in which the many-body interaction between the molecular species may generally be decomposed into convergent pair-, triplet-, ... interactions.

Finally, we turn to the discussion of the periodicity of the persistent current.  On the basis of their extensive numerical simulations for a weakly disordered 10-channels ring Bouchiat and Montambaux\cite{9} have found that the persistent current averaged over a range of electron numbers has periodicity $\phi_0/2$.  However, the physical origin of this period halving remained  unclear\cite{18}.  As shown by \eqref{eq50} the analysis of \cite{8} leads to period halving of the persistent current in a disordered one-dimensional ring while the discussion below for the two-channel ring indicates that besides the periodicity $\phi_0/2$ there exists another approximate periodicity, $\phi_0$, which  is induced by the disorder.

As mentioned earlier we consider the typical case of a two-channel ring where the lower energy-band (\ref{eq28b}, \ref{eq31b}) is completely filled and the upper band (\ref{eq28a},\ref{eq31a}) is half-filled, with $N_e=N/2$ electrons.  The filled lower band carries no free particle current (vanishing fermi velocity) and the total current in the upper band is approximately given by

\begin{equation}\label{eq51}
I\simeq -I_0\biggl(\frac{2\phi}{\phi_0}-\frac{1}{2}\biggr)
-\frac{\varepsilon^2_0\pi e}{2h}
\sin \frac{2\pi \phi}{\phi_0}
\biggl(2+\frac{1}{\cos^2\frac{2\pi\phi}{\phi_0}}\biggr),
0<\frac{\phi}{\phi_0}<\frac{1}{2}
\quad .
\end{equation}		
The first term is the free particle current \eqref{eq49} averaged over even and odd numbers of electrons in the upper band.
The second term determines the dominant decrease of the persistent current $I^{(2)}=\sum_{k,k'}I^{(2)}_{s_1s_2}(k,k',\phi)$ in the two-channel ring which is induced by the disorder.  It is given by \eqref{eq46} by ignoring the weak perturbation arising from summation of $f_{s_1s_2}(k,k')$ over the occupied individual energy levels averaged over electron numbers, whose magnitude is typically of order 1 at $\phi\sim\phi_0/4$, relative to the larger effect of the dominant flux dependence in \eqref{eq51} around this flux value.  This result for the order of magnitude of $\sum_{k,k'}f_{s_1s_2}(k,k')$ is inferred from the magnitude of the corresponding simpler sum, $\sum_k f_s (k)=\sum 1/\sin s, s=\frac{2\pi}{N}\biggl(k+\frac{\phi}{\phi_0}\biggr), k=1,2,\ldots N$, which arises in the study of the effect of disorder on persistent current in a one-dimensional ring \cite{8} and whose approximate evaluation yields results linear in $\phi/\phi_0$ both for even and for odd numbers of electrons, for $0<\phi/\phi_0 < 1/2$.  Finally we note that the above estimate of the persistent current near $\phi=\phi_0/4$, which focuses on finding its order of magnitude, does not necessarily yield the correct sign of the current.

The existence of two distinct components of persistent current, a free particle component of periodicity $\phi_0/2$ and a component of periodicity $\phi_0$ due to the disorder\cite{19} may reflect the presence of two distinct periods in the tight-binding equations (\ref{eq3}-\ref{eq8}) (period $N\phi_0$) and in the boundary condition \eqref{eq1} (period $\phi_0$).  Note that in the single-channel (one-dimensional) case the disorder effect itself in the persistent current before averaging over particle numbers involves the periodicity $\phi_0/2$ (see Eqs. (49-50) of \cite{8}).

\section{CONCLUDING REMARKS}\label{secV}

We conclude with a general remark concerning the possibility of interpreting further specific features of the numerical results for persistent current averaged over numbers of electrons in many-channel disordered systems in fig. 4 of Bouchiat and Montambaux\cite{9} in terms of disorder effects in few-channel systems, say $M=3\;\text{or}\;4$.

The definition of independent channels in a non-disordered $M$-chain tight-binding cylinder threaded by a flux $\phi$ involves in a first step the diagonalization of the interchain hopping, $t$, leading to the energy eigenvalues\cite{20}

\begin{equation}\label{eq52}
E_n=-2t\cos\frac{n\pi}{M+1}-2t\cos
\biggl[\frac{2\pi}{N}(k_n-\phi)\biggr]\quad ,
\end{equation}•		
for the channels $n=1,2,\ldots M$, where $k_n=0,\pm 1,\pm 2\ldots$.  For $M=2$ this expression reduces clearly to \eqref{eq28} with the definition \eqref{eq31} of wavenumbers.  However, for larger $M$ the discussion of the effect of disorder becomes rapidly very complicated analytically.  Nevertheless, specific features in the persistent current, such as the existence of weaker peaks and/or dips at special flux values, which are related to the interchain hopping may be expected in analogy to what we have found for the single-chain- and for the two-chain cylindrical ring model studied above.


\begin{thebibliography}{unsrt}
\bibitem{10} The necessity of averaging over even and odd numbers of electrons is due to the extreme sensitivity of the persistent current to the numbers of electrons in the rings, which may vary from ring to ring. This has been discussed in \cite{9} for multichannel rings and also earlier for one-dimensional rings in Ref. \cite{11}.
\bibitem{16} The analog  of the fixed overall flux-dependent factor (of period $\phi_0$) in the persistent current of the $k,k'$-level pair in \eqref{eq46} is the factor $\sin^{-2}(2\pi\phi/\phi_0$) (of period $\phi_0/2$) in the persistent current in a level $k$ in the one-dimensional case (Eq. (38) of \cite{8}).
\bibitem{17} The relation between the rectangular distribution of width $W$ for the site energies used in Ref. \cite{9} and the Gaussian parameter $\varepsilon_0$ is obtained by identifying the second moment of the rectangular distribution with $\varepsilon^2_0$, which yields $\varepsilon^2_0=W^2/12$.
\bibitem{19}The sine-function periodicity $\phi_0$ in the dominant disorder effect (near $\phi=\frac{\phi_0}{4}$) in the persistent current in the two-channel case is to be compared with the periodicity $\frac{\phi_0}{2}$ of the corresponding dominant disorder effect \eqref{eq50} near $\phi=0$ and near $\phi_0/2$ in the single-channel case (which in that case leads to a periodicity $\phi_0/2$ in the current after particle number averaging).  This difference between the flux dependences of the dominant disorder effect in the persistent current in the single- and two-channel cases is the origin of the existence of two dominant periodicities of the persistent current in the two-channel case, Eq. \eqref{eq51}, which would not be affected significantly by averaging over particle numbers in (\ref{eq46},46a).
\end{thebibliography}
\end{document}